\documentclass[pra,twocolumn,superscriptaddress,showpacs,amssymb]{revtex4}

\usepackage{epsfig} 
\usepackage{graphics}

\newcommand{\be}{\begin{equation}}
\newcommand{\ee}{\end{equation}}
\newcommand{\bea}{\begin{eqnarray}}
\newcommand{\eea}{\end{eqnarray}}
\newcommand{\ket}[1]{\left | #1 \right \rangle}
\newcommand{\bra}[1]{\left \langle #1 \right |}

\begin{document}

\title{
Entanglement of two atomic samples by quantum non-demolition
measurements.}

\author{Antonio Di Lisi} 
\email[E-mail:]{dilisi@sa.infn.it}
\affiliation{Dipartimento di Fisica ``E. R. Caianiello'', Universit\`a di Salerno,
INFM--Unit\`a di Salerno, I-84081 Baronissi (SA),  Italy}
\author{Klaus M{\o}lmer}
\email[E-mail:]{moelmer@phys.au.dk}
\affiliation {QUANTOP, Department of Physics and Astronomy,
University of {\AA}rhus, DK 8000 {\AA}rhus C., Denmark}


\begin{abstract}
This paper presents simulations of the state vector dynamics for
a pair of  atomic samples which are being probed 
by phase shift measurements on an optical beam passing through
both samples. We show how measurements, which are sensitive 
to different atomic components, serve to prepare states which
are close to being maximally entangled.
\end{abstract}


\maketitle

\section{Introduction}\label{section_1}

The population of an atomic  state $|1\rangle$ 
can be probed by a phase shift measurement on a field
that couples $|1\rangle$ non-resonantly to another
atomic state.
This implies that for an atomic sample with all atoms
populating two long lived states $|1\rangle$ and $|2\rangle$,
it is possible to count non-destructively the number $n_1$ of
atoms in $|1\rangle$. The state vector or density matrix
of the sample with, e.g.,  an initial  binomial distribution of populations 
$n_1$ will be modified, as the quantum mechanical uncertainty of the
number $n_1$ is reduced during the measurement 
\cite{kuz98,moelmer99,bouchoule02}.
This effect has been demonstrated 
experimentally \cite{kuz99,Kuz00}, and further theoretical
proposals \cite{Duan00} have addressed the possibility to 
entangle pairs of samples by performing joint phase shift 
measurements, where light is propagated through both samples
and thus provides information about total occupancies of various
states.  In particular it has been  possible to prepare 
EPR-correlated samples \cite{juls01} where the entanglement could be 
experimentally proven by a quantitative test  involving spin noise
measurements \cite{ciracgauss,simongauss}.

The theoretical analysis in \cite{Duan00,juls01} addresses
the behavior of collective atomic spin operators due  to their
interaction with the field operators (in the Heisenberg picture).
Following the analysis in \cite{bouchoule02}, 
in the present paper we present wave function simulations, in which
we display how the sequence of photo-detection events gradually modifies
the state vector of the samples. The two approaches are of course
equivalent, but they may provide different insights, and in addition
the present state vector approach does not rely on the atomic operators 
being approximated by harmonic oscillators, i.e., we can treat cases
where the mean polarization changes significantly in the samples, which
is not practically possible in the operator formulation.

The paper is organized as follows. 
In Sec. \ref{detection model}, we introduce our physical model for the measurement
of optical phase shifts, and we present the formal description
of the state vector evolution conditioned on the photo-detection
record.
In Sec. \ref{numerical}, we show results of simulations where the atoms are first
all prepared in superposition states of the two internal states
prior to detection of the number of atoms in state $|1\rangle$.
We show that the resulting state vector is entangled, and we show that a
subsequent spin rotation of all atoms followed by new measurements will lead
to stronger entanglement. 
In Sec. \ref{measurements}, we show simulations where the atomic samples are subject
to  continuous spin rotation during measurement, corresponding to the
experimental situation of Ref.\cite{juls01}. In  Sec. \ref{discussion}, we present
an analysis and interpretation of the results, and 
Sec. \ref{conclusion} concludes the paper.

\section{Optical phase shift measurements}\label{detection model}

\begin{figure}[htbp]
\begin{center}
\includegraphics[height=4cm]{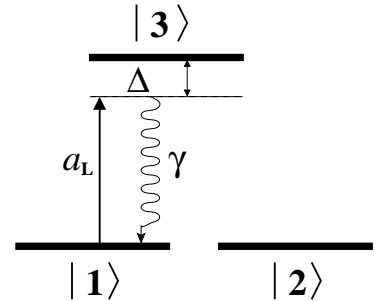}
\end{center} 
\caption{ 
Level structure of the atoms. The states $\ket{1}$ and $\ket{2}$ are
stable states, $|1\rangle$ is coupled off-resonantly by the probe to 
the excited state $\ket{3}$.}
\label{ss}
\end{figure}

In our model we are considering clouds of atoms whose relevant level 
structure consists of three levels: two degenerate ground states,
$\ket{1}$ and $ \ket{2}$, and an exited state, $\ket{3}$. 
We suppose that the only permitted transition is 
$\ket{1}\rightarrow\ket{3}$, excited
by the laser beam passing through the cloud and
followed by spontaneous decay with emission rate $\gamma$, cf.,
Fig.~\ref{ss}.

\begin{figure}[htbp]
\begin{center}
\includegraphics[width=8.5cm]{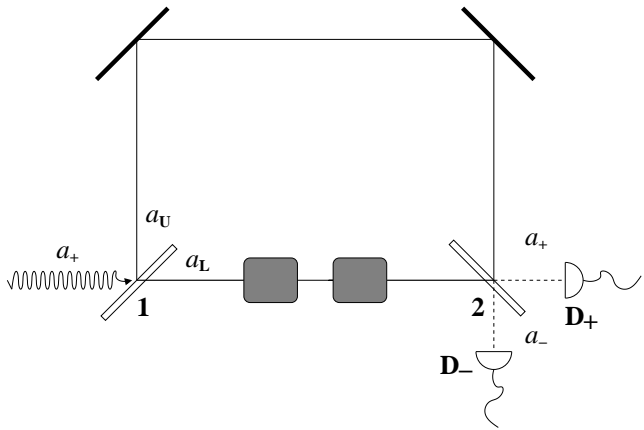}
\end{center} 
\caption{ 
Atoms occupying the internal state $|1\rangle$ in the two samples interact 
with the light field which is  incident from the left in the figure. 
The phase shift of the light field due to interaction with these  atoms 
is registered by the different photo-currents in the two detectors.}
\label{es}
\end{figure}

Our scheme is based on the detection of 
photons passing through the clouds. The scheme only requires 
classical input fields, since the state reduction
following every single photodetection event, {\it a posteriori},
extracts the effect of the interaction of a single photon with 
the atomic samples.  
There are different possibilities to create a detection
scheme that measures the population of a single atomic state, or the
population difference between two atomic states. One can use light 
with two polarization components that interact differently with the
two internal states, and hence the different phase shifts lead to
a polarization rotation signal; or, one can use frequency modulated
light, where one frequency component is closer to resonance than the
other, and hence a phase difference between the two results from the
interaction between the atoms. For ease of presentation, we follow
the description in \cite{bouchoule02}, and imagine an interferometric
setup, where a field is split in two paths, one that interacts with the
atomic samples and one that propagates through free space. 
This experimental set-up is schematically shown in Fig.~\ref{es}. A phase
difference between these two fields due to the atoms can be
resolved by the intensities measured by
photo-detectors $D_{+}$ and $D_{-}$ in the two output ports of the
interferometer. 
We may enclose the atomic samples in optical cavities
to enhance the interaction with the atoms by
passing the light through each sample many times. This will
also reduce the effect of spontaneous scattering which will
hence be omitted from the present analysis. As shown in 
\cite{bouchoule02} wave function simulations can incorporate
spontaneous scattering, and it can also be estimated in the 
operator formulation \cite{Duan00}.

Let us introduce the atomic transition operators
$\sigma_{ij}^{n}=\ket{i}_{n}\!\bra{j}$, where $i(j)=1,2,3$
and $n$ enumerates the atoms.
In the dipole and rotating wave approximation, the interaction
Hamiltonian is:
\begin{equation} \label{hamiltonian}
H_{I} = \sum_{n}g(\sigma_{13}^{n}a_{L}^{\dag} + \sigma_{31}^{n}a_{L})
\end{equation}  
where $g$ is the atom-photon interaction strength, and $a_{L}$ is the
annihilation operator for the field component interacting with the atoms.
  
If $\Delta$ is the detuning between the light field and the atomic
transition frequency $(E_{3}-E_{1})/\hbar$, the second order transition 
amplitude for the interaction process is:
\begin{equation}
\mathcal{I}_{fi} =
\frac{g^{2}\Delta}{\Delta^{2}-\frac{\gamma^2}{4}}
-i\frac{g^{2}\frac{\gamma}{2}}{\Delta^{2} -\frac{\gamma^{2}}{4}}
\end{equation}
where the indices $i$ and $f$ represent the initial and final state of
the process, which  in our case is the same state of an atom in
$|1\rangle$ and one photon of the radiation field.

In the case of off resonant scattering, with $\Delta\gg\gamma$, we get 
\begin{equation}
\mathcal{I}_{fi} =
\frac{g^{2}}{\Delta}.
\end{equation}  
which is just the energy shift of state $\ket{1}$
due to the absorbed and re-emitted photon. Hence, the time evolution 
of the atom-photon state assumes the form 
$\ket{1}\otimes\ket{n_L=1}\stackrel{\tau}{\longrightarrow}
e^{-i\frac{g^2}{\Delta}\tau}\ket{1}\otimes\ket{n_L=1}$.

Defining $\chi = \frac{g^2}{\Delta} $, we can write an effective
Hamiltonian describing the interaction between a single mode
light field and a gas of $N$ atoms:
\begin{equation}\label{eff.hamiltonian}
H_{eff} \equiv \sum_{n}\chi\ket{1}_{n}\!\bra{1}a_{L}^{\dag}a_{L} = 
\sum_{n}\chi \sigma_{11}^{n}a_{L}^{\dag}a_{L}.
\end{equation}  

By introducing the atomic spin operators for the ground states
\begin{eqnarray}
j_{nz} & = &
\frac{1}{2}\left(\sigma_{22}^{n} - \sigma_{11}^{n}\right) \\
j_{n+} & = & \frac{1}{2}\sigma_{21}^{n}\\
j_{n-} & = & \frac{1}{2}\sigma_{12}^{n},
\end{eqnarray} 
the effective Hamiltonian (\ref{eff.hamiltonian}) becomes
\begin{equation}\label{eff.hamiltonian2}
H_{eff} = \sum_{n=1}^{N}\chi\left(\frac{1}{2} -j_{nz}\right)a_{L}^{\dag}a_{L}.
\end{equation}

In an ensemble of atoms, where each atom is initially prepared in the
same state, and where the interaction with the surroundings is
identically the same for all atoms, the collective atomic state
retains the full permutation symmetry, and it is convenient to expand 
this collective state on eigenstates for the effective collective
angular momentum: 
\begin{equation}\label{1wavefunction}
\ket{\Psi}= \sum_{M=-J}^{J}\mathcal{A}_{M}\ket{J,M},
\end{equation}
where $J=\frac{N}{2}$ is the total angular momentum, and 
$M = \frac{1}{2}\left(n_{2}-n_{1}\right)$ is the eigenvalue of
the operator $J_{z}= \sum_{n}^{N} j_{nz}$. Collective raising and
lowering operators, and the corresponding cartesian $x$- and $y$-
components of the collective angular momentum, are defined 
as similar sums over all atoms in the sample.

A photon incident on the interferometer splits as a superposition
of a photon in the upper path and a photon in the lower path,
$\frac{1}{\sqrt{2}}(\ket{U}_{ph}+\ket{L}_{ph})$, and the
time evolution of the collective atomic state due to the interaction
with this photon for a duration $\tau$ is given by
\begin{widetext}
\begin{eqnarray}
\ket{\Phi(\tau)}_{ph+at}&=&
e^{-i\sum_{n=1}^{N}\chi\left(\frac{1}{2} -j_{nz}\right)a_{L}^{\dag}a_{L}
\tau}\frac{1}{\sqrt{2}}(\ket{U}_{ph}+\ket{L}_{ph})\otimes\ket{\Psi}\nonumber \\
&=&\sum_{M=-J}^{J}
\frac{\mathcal{A}_{M}}{\sqrt{2}}\left(\ket{U}_{ph}+
e^{-i\left(\frac{N}{2}-M\right)\chi\tau}\ket{L}_{ph}\right)
\otimes\ket{J,M},
\end{eqnarray}
\end{widetext}          
where $\left(\frac{N}{2} - M\right) = \left(\frac{n_{1} + 
n_{2}}{2} - \frac{n_2 -
n_1}{2}\right) = n_1$. Hence, as claimed above, the phase
shift of the single photon state $|L\rangle_{ph}$  is proportional 
to the total population of state $\ket{1}$.

\subsection{Entanglement due to measurement}

In \cite{bouchoule02} the interaction with a single sample is analyzed
in detail, and the possibility to achieve spin squeezing due to 
the interaction with the atoms is analyzed. We now turn our attention
to a system of two atomic ensembles of $N_1$ and $N_2$ atoms, respectively. 
Within each sample we assume permutation symmetry among the atoms, {\it
i.e.}, each sample is represented by the collective spin states,
introduced above.

The state vector of the samples can be expanded on product state wave
functions of the two ensembles
\begin{equation}
\ket{\Psi} = 
\sum_{M_{1},M_{2}}\mathcal{A}_{M_1,M_2}\ket{M_{1},M_{2}},
\end{equation}
where $|M_1,M_2\rangle \equiv 
|J_1=N_1/2,M_1\rangle\otimes|J_2=N_2/2,M_2\rangle$, and where 
we assume an initial product state $\mathcal{A}_{M_1,M_2} =
\mathcal{A}_{M_1}\mathcal{A}_{M_2}$ before we submit
the system to the incident field.
The photon interacts with each of the ensembles in the lower
arm of the interferometer, and the state of the system
after the interaction, but before the
photo-detection, is
\begin{widetext}
\be
\ket{\Phi(\tau)}_{ph+at}=
\sum_{M_1,M_2}\mathcal{A}_{M_1,M_2}\left(\frac{\ket{U}_{ph}
+e^{-i\left[\frac{N_1 + N_2}{2}-\left(M_1
+ M_2\right)\right]\chi\tau}\ket{L}_{ph}}{\sqrt{2}}\right)\otimes\ket{M_1,M_2}
\ee
\end{widetext}
The detectors $D_{+}$ and $D_{-}$ click, when they detect a 
photon in one of the states $\ket{\pm}_{ph} =
\frac{1}{\sqrt{2}}\left(\ket{U}_{ph}\pm\ket{L}_{ph}\right)$,
and associated with this detection event the photon is destroyed by the
action of one of the annihilation operators $a_{+}$ and $a_{-}$.
It is thus useful to rewrite the photon-atom  state  explicitly in
terms of the two components distinguished by the photodetectors:
\begin{widetext}
\bea
\ket{\Phi(\tau)}_{ph+at}&=&
\sum_{M_1,M_2}\mathcal{A}_{M_1,M_2}
\left(\frac{1+e^{-i\left[\frac{N_1 + N_2}{2}-\left(M_1
+ M_2\right)\right]\chi\tau}}{2}\right)\ket{M_1,M_2}\otimes\ket{+}_{ph} \nonumber\\
& &+\sum_{M_1,M_2}\mathcal{A}_{M_1,M_2}
\left(\frac{1-e^{-i\left[\frac{N_1 + N_2}{2}-\left(M_1
+ M_2\right)\right]\chi\tau}}{2}\right)\ket{M_1,M_2}\otimes\ket{-}_{ph}.
\eea 
\end{widetext}
The probabilities that the  photon is detected in state $\ket{+}_{ph}$ or in
state $\ket{-}_{ph}$, are
\begin{center}
\bea
\pi_{+} &=&
\sum_{M_1,M_2}\left|\mathcal{A}_{M_1,M_2}\right|^2 \nonumber\\
&\times&\cos^{2}{\left[\left(\frac{N_1+N_2}{4}-\frac{M_1+M_2}{2}\right)\chi\tau\right]},\label{+prob}\\
\pi_{-} &=&
\sum_{M_1,M_2}\left|\mathcal{A}_{M_1,M_2}\right|^2 \nonumber \\
&\times&\sin^{2}{\left[\left(\frac{N_1+N_2}{4}-\frac{M_1+M_2}{2}\right)\chi\tau\right]}.\label{-prob}
\eea 
\end{center} 

Thus, after the detection of a photon, the state of the two ensembles
of atoms is projected, with probability $\pi_{+}$ and $\pi_{-}$
respectively, into one of the following states
\bea
\ket{\Psi}_{+}&=&\frac{1}{\sqrt{C_+}}\sum_{M_{1},M_{2}}\frac{\mathcal{A}_{M_1,M_2}}{2}\nonumber \\
&\times&\left(1+e^{-i\left[\frac{N_1+N_2}{2}-\left(M_1+M_2\right)\right]\chi\tau}\right)
\ket{M_{1},M_{2}}\label{+state}\\
\ket{\Psi}_{-}&=&\frac{1}{\sqrt{C_-}}\sum_{M_{1},M_{2}}\frac{\mathcal{A}_{M_1,M_2}}{2}\nonumber \\
&\times&\left(1-e^{-i\left[\frac{N_1+N_2}{2}-\left(M_1+M_2\right)\right]\chi\tau}\right)
\ket{M_{1},M_{2}}\!,\label{-state}
\eea
where $C_{\pm}$ are normalization factors. From Eqs. (\ref{+state})
and (\ref{-state}) we observe the entanglement between
the two atomic ensembles emerge.   

The detection procedure, and the corresponding  wave function updating,
is repeated for a number $N_{ph}$ of photons, 
being detected one at a time. Hence, by defining the following
\emph{entangling} factors  
\begin{equation}
\mathcal{F}_{\pm}(M_1+M_2) = 
\left(\frac{1 \pm e^{-i\left[\frac{N_1+N_2}{2}-\left(M_1+M_2\right)\right]\chi\tau}}{2}\right),   
\end{equation}
after $N_{ph}$ detected photons, of which $N_+$ are  detected by $D_{+}$ and
$N_-=N_{ph} - N_+$ are detected by $D_{-}$, the state vector of the samples 
is
\bea \label{npstate}
\ket{\Psi}_{N_{ph}}&=& \frac{1}{\sqrt{C}}\sum_{M_{1},M_{2}}\mathcal{A}_{M_1}\mathcal{A}_{M_2}
\left[\mathcal{F}_{+}(M_1+M_2)\right]^{N_+}\nonumber \\
& &\times\left[\mathcal{F}_{-}(M_1+M_2)\right]^{(N_{ph}-N_+)}
\ket{M_{1},M_{2}}.
\eea
%

The square norm of the entangling factor in equation (\ref{npstate})
is 
\bea
\mathcal{B}(M_{12}) &=& 
\left[\cos\left(\frac{N-M_{12}}{2}\chi\tau\right)\right]^{2N_+}\nonumber\\
& &\times\left[\sin\left(\frac{N-M_{12}}{2}\chi\tau\right)\right]^{2(N_{ph}-N_+)},
\eea
where $M_{12}= M_1 + M_2$, and where, for simplicity, we assume the same
number of atoms in each sample $N=N_1=N_2$.  
For large $N_{ph}$ this function is very
peaked with maxima at values $M_{12} = \overline{M}$
obeying
\be \label{tan}
\tan\left(\frac{N-\overline{M}}{2}\right)\chi\tau =
\pm\sqrt{\frac{N_{ph}-N_+}{N_+}}.
\ee
Equation (\ref{tan}) has multiple solutions for $\overline{M}$, due to
both the periodicity of the $tan$-function and the double sign in the right
hand side, but we can remove this ambiguity in the solutions 
if we ensure that only a single maximum is compatible with the
initial distribution of the two samples in $M_1$ and $M_2$.
Calculating the second derivative of $\mathcal{B}(M_{12})$  around the maximum
values, we get an estimate of the width of the peak in
$M_{12}$
\be \label{secder}
\left.\partial_{M_{12}}^{2}\mathcal{B}(M_{12})\right|_{\tiny{M_{12}=\overline{M}}}=
-(\chi\tau)^{2}N_{ph}\mathcal{B}(M_{12}).
\ee
If the number of atoms is large enough, we can approximate
$\mathcal{B}(M_{12})$ with a Gaussian, and with this approximation
we obtain the r.m.s width 
\be \label{gaus.var}
\Delta M_{12}=\frac{1}{\chi\tau\sqrt{N_{ph}}}.
\ee  
Thus, the more photons that are detected, the more peaked is the
distribution in $M_{12}$.

\section{Numerical simulation, consecutive measurements of
$J_{z1}+J_{z2}$ and $J_{y1}-J_{y2}$}\label{numerical}

In this section, 
the detection model described in the previous section will be 
implemented in a numerical simulation in which the two samples 
have the same, not very large, number $N$ of atoms and
the initial state is the eigenstate of the operator
$J_{x1}+J_{x2}$ with eigenvalue $J_1+J_2 = 2J =N$, {\it i.e.},
all the atoms are initially prepared in the superposition
state $\frac{1}{\sqrt{2}}(|1\rangle+|2\rangle)$.
The corresponding amplitudes in the basis eigenstates of the 
$z$-component of the collective angular momentum operators
$|J,M_i\rangle$ are
\be\label{iamp}
\mathcal{A}_{M_i} =
\left(\frac{1}{2}\right)^{J}\sqrt{\frac{(2J)!}{(J+M_{i})!(J-M_{i})!}}
\; ,
\ee
where $i=1,2$. Expressed in terms of the number of atoms in state
$|1\rangle$ in each sample, $n_{1}^{i}$, the square of the 
amplitude (\ref{iamp}) is
$|\mathcal{A}_{M_i}|^2 =\left(\frac{1}{2}\right)^{N}\frac{N!}{(N-
n_{1}^{i})!(n_{1}^{i})!}$. This means that
in each ensemble the atoms are distributed in state $\ket{1}$ and
$\ket{2}$ according to a binomial distribution with probability
$\frac{1}{2}$.

\begin{figure}[htbp]
\begin{center}
\includegraphics[width=8.5cm]{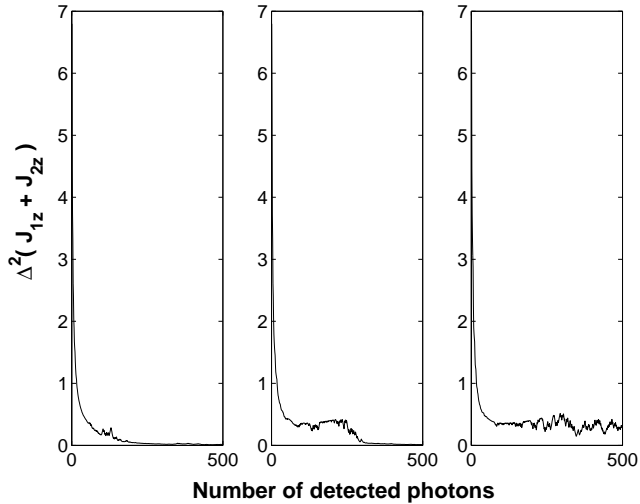}
\end{center}
\caption{ 
Variance of $J_{1z}+J_{2z}$ as function of the number of detected
photons. Each sample consists of 20 atoms in the simulations, and 
the phase angle $\chi\tau$ per atom and photon is
0.24 in the simulation.}
\label{var}
\end{figure}

Beginning
with an initial state characterized by amplitudes (\ref{iamp}), 
we proceed with  a series of photo detections that project
the state of the system according to Eqs. (\ref{+state}) and
(\ref{-state}). After a large
number of photons $N_{ph}$ has been detected, the uncertainty
in $M_{12}$ is reduced, {\it i.e.}, we are almost in one of 
the eigenstates of
$J_{1z}+J_{2z}$. Fig.~\ref{var} shows  
the behavior of the variance $\Delta^2(J_{1z}+J_{2z})$ 
as a function of the number of detected photons
for three different simulation records.
It is evident that when the
number of detected photons is large  enough,
$\Delta^2(J_{1z}+J_{2z})$ goes to zero, i.e we have a definite  
value of $\overline{M}= M_1+M_2$. In the simulations we assume
a phase angle $\chi\tau=0.24$. Such a large phase shift on the atomic
state due to interaction with a single photon is not realistic in
experiments with freely propagating fields. In our simulations
a  smaller value of $\chi\tau$ just
implies that more photons have to be detected to achieve the same 
reduction in $\Delta^2(J_{1z}+J_{2z})$, c.f., (\ref{gaus.var}).

To quantify the entanglement between the two samples we apply the
definition by Bennett et al \cite{bennett96} of the entanglement
between two systems in a pure state $|\Psi\rangle$:
\be
\mathcal{E} = Tr(\rho_1 \log_2\rho_1)=Tr(\rho_2 \log_2\rho_2),
\ee
where $\rho_{1}=Tr_{2}\ket{\Psi}\bra{\Psi}$ is the reduced density
matrix of system 1. (Similarly for $\rho_2$). If $N$ is the number of
atoms in each sample, and we restrict ourselves to states which are
symmetric under permutations inside samples, the quantity
$\mathcal{E}$ takes values between zero for a product state, and
$\log_2(N+1)$ for a maximally entangled state of the samples. 

\begin{figure}[htbp]
\begin{center}
\includegraphics[width=8.5cm]{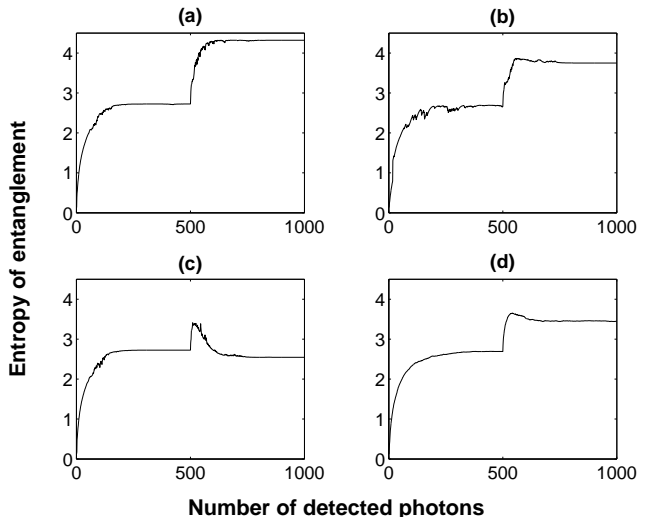}
\end{center}
\caption{ 
Entanglement of two samples of each $N=20$ atoms. After detection of the first
$N_{ph}=500$ photons, the samples are rotated $\pm 90$ degrees in spin space,
and the subsequent detection events lead, typically, 
to further entanglement. Figs.~4~(a-c) show results of three different
simulation records, Fig.~4(d) shows the average over 50 such simulations.
$\chi\tau = 0.24$ as in Fig. 3.}
\label{ent1}
\end{figure}

Fig.~\ref{ent1} shows the behavior of the entropy in three different
simulations. The left hand side of each figure, for $N_{ph} < 500$
corresponds to the detection, analyzed above. To prove that the samples
are entangled and not just classically correlated, in the experiments
of \cite{juls01} a second observable was introduced, for which the
quantum mechanical uncertainty was similarly reduced, and 
measurements on both  sets of observables conclusively demonstrate the
entanglement \cite{ciracgauss,simongauss}. In our theoretical calculation,
we know of course already that the state is entangled,
but as we shall see, we can increase the
entanglement by taking a second round of measurements.

Thus, after the detection of 500 photons, we apply opposite
rotations to the atomic samples, and we proceed with 
similar measurements as before, which effectively measure
$J_{1y}-J_{2y}$ in the non-rotated frame.
After the rotations, the full permutation symmetry of the system is broken, 
i.e the total angular momentum $J^2 = J_{x}^2+J_{y}^2+J_{z}^2$ is not
conserved. The final state of the two samples is still an
eigenstate of both $J_{1}^2$ and $J_{2}^2$, but with
different values of the total $J$. Our formalism (25,26) is already
prepared to handle that situation, since the basis vectors in the
expansion are kept on the form of product states rather than angular
momentum coupled states of the two samples. 

As illustrated by the right hand parts of the panels in
Fig.~\ref{ent1}, the entanglement had essentially saturated after
the first 500 detection events, but after the rotation, it grows again
to saturate, typically, at a higher level.
In Fig.~\ref{ent1}(a) we show an example of the  evolution of the entropy
in which the final value ($\mathcal{E}_{N_{ph}}= 4.3193$) 
is very close to that of the maximally entangled state
($\log_{2}(20+1)=4.3923$). In Fig.~\ref{ent1}(c), however, we show a 
case in which the final entropy value is reduced by the second set of
rotations. In Sec. \ref{discussion}, we shall return to an analysis of these
findings. Finally,  Fig.~\ref{ent1}(d) presents the average of the
entropy over 50 simulations. 
           
\section{Measurements with continuous rotation of spin
components}\label{measurements}

As we just saw that measurements of two different sets of operators
typically lead to an increase of the amount of entanglement, it is natural
to consider a continuous change between these operators.
In fact, in the experimental work by Julsgaard et al \cite{juls01},
such a continuous rotation was induced for purely practical reasons
(so that the relevant signal could be picked up in the absence of
technical noise at a high frequency).
In this section we present the results of simulations in which 
continuous opposite rotations are applied to the atomic spins of the
two samples  as the photo detection proceeds. 
If $\mathcal{A}_{M_1M_2}^{N_{ph}}$ is the wave function amplitude of the
system after $N_{ph}$ photo detections, the updated state
vector rotated  by the angle $\pm \theta$ prior to the subsequent detection is
\begin{widetext}
\be
\ket{\Psi}_{N_{ph}} = \sum_{M^{'}_{1},M^{'}_{2}}
\sum_{M_{1},M{2}}\mathcal{D}^{J_1}_{M_1M^{'}_1}\left(\theta\right)
\mathcal{D}^{J_2}_{M_2M^{'}_2}\left(-\theta\right)
\mathcal{A}_{M_1M_2}^{N_{ph}}
\ket{M^{'}_{1},M^{'}_{2}},
\ee
\end{widetext}   
where  
$D^{J_i}_{M_iM^{'}_i}\left(\theta\right)=
\bra{M_i,J_i}e^{-iJ_{ix}(\theta)}\ket{J_i,M_{i}^{'}}$
is the rotation matrix element for sample $i=1,2$. 
The wave function
updating algorithm has the same structure as described in section
\ref{detection model}.  In a realistic
experimental situation, one would detect photons according to
a Poisson process, and with a constant rotation frequency, induced,
{\it e.g.}, by application of opposite DC magnetic fields onto the atoms,
this would lead to small rotation angles with an exponential
distribution law. For simplicity, however, we apply the same
small rotation angle $\theta$ between each detection event.
 
\begin{figure}[htbp]
\begin{center}
\includegraphics[width=8.5cm]{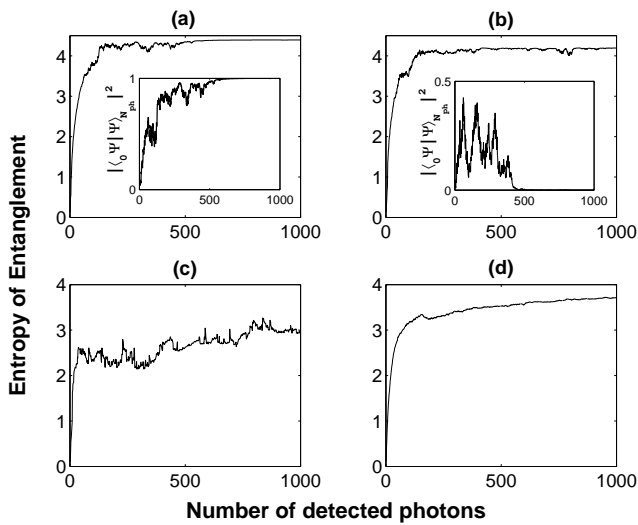}
\end{center}
\caption{ 
Entanglement of atomic samples  with 20 atoms. The spins are
rotated in opposite directions around the $x$-axis
by the angle $\theta=\frac{\pi}{5}$ after each detection event.
Insets (a) and (b) show the evolution of the overlap
$|\bra{\Psi_0}\Psi\rangle_{N_{ph}}|^2$ between the state of the 
samples and the maximally entangled state (See section
{\upshape(\ref{discussion})}). In (a) $\ket{\Psi}_{N_{ph}}$ converges
towards $\ket{\Psi_0}$ and further detections have no effect on the
state vector. In (b) the state gradually gets rid of its component
along $\ket{\Psi_0}$, and it subsequently evolves in the orthogonal 
subspace of $\ket{\Psi_0}$. In (c) is shown a record where the
entanglement is smaller than in the simulations in (a) and (b),
and in (d) is shown the average over 50 simulations.}
\label{ent2}
\end{figure}

In Fig.~\ref{ent2} we show realizations of the detection scheme with
small rotations. The rotation angle between subsequent photodetection
events is $\theta=\frac{\pi}{5}$.
Fig.~\ref{ent2}(a) shows a case where the entropy evolves to the maximum
value: $\mathcal{E}_{N_{ph}} = 4.3923$.  
In Fig.~\ref{ent2}(b) and \ref{ent2}(c) we present 'typical' and  
'worst case' results of the simulations, while in Fig.~\ref{ent2}(d) 
we provide the average over 50 simulations. 

We observe in these figure that compared to the results shown in the 
previous section, there is a faster
increase of the entanglement towards an almost constant level,
which varies from one simulation to the other.

\section{Analysis and interpretation}\label{discussion}

In order to interpret the results obtained in the
previous sections,  let us first address the achievements 
of probing two operators rather than a single one. The improvement
is quite easily understood, if one notes that the detection of
$J_{z1}+J_{z2}$ effectively produces an approximate eigenstate
of this operator. This state has amplitudes on different 
$M_1$ and $M_2$ states with $M_1+M_2$ fixed by the measured value, 
but the eigenspace is degenerate, and the amplitudes are
simply proportional to the ones in the initial state.
Due to the initial binomial distribution on $M_1$ and $M_2$,
the distribution over, {\it e.g.,} $M_1$ will therefore have a width
of approximately $\sqrt{N}$. The reduced density matrix has the 
corresponding number of non-vanishing
populations, suggestive of $\mathcal{E} \sim \log_{2}\sqrt{N}
=0.5 \log_{2}N$, half of the maximal value. This argument
accounts for the first plateau reached in Fig. 4. 

The measurement
of the $z$-components causes a broadening of the distribution
on $J_y$ eigenstates beyond the initial distribution, which was
also binomial in that basis. The subsequent measurement
of $J_{y1}-J_{y2}$ will produce a state with $M_{1y}-M_{2y}$
fixed by the measurement, but within the degenerate space of 
states with this fixed value the distribution on $M_{1y}$ of
the reduced density matrix is broader than $\sqrt{N}$, and the 
entanglement is correspondingly larger. This accounts for the increase
of entanglement seen in most simulations.

For large collections of atoms with large mean values of the collective 
$J_x$ operators,
the orthogonal spin components $J_z$ and $J_y$ are well approximated
by effective position and momentum operators, and for two particles,
the pair of combinations of position and momentum
operators $x_1+x_2$ and $p_1-p_2$ commute, {\it i.e.}, they can both be
measured with high precision. But, this is not an 
exact replacement, and in our simulations with fewer atoms, we see
significant deviations from this picture. The commutator of
$J_{z1}+J_{z2}$ and $J_{y1}-J_{y2}$ is proportional to the
operator $J_{x1}-J_{x2}$ which does not vanish, and in general,
measurements sensitive to one of these operators are complimentary
to measurements sensitive to the other one.  
It is possible, however, to find {\it a single} joint eigenstate of the
operators \cite{wang02}:\\
\begin{eqnarray}
(J_{1x}-J_{2x})|\Psi_0 \rangle &=&0, \label{eq:jj1}\\
(J_{1y}-J_{2y})|\Psi_0 \rangle &=&0. \label{eq:jj2}
\end{eqnarray}
Rewriting these equations as 
\begin{eqnarray}
(J_{1+}-J_{2+})|\Psi_0 \rangle &=&0, \\
(J_{1-}-J_{2-})|\Psi_0 \rangle &=&0,
\end{eqnarray}
it is easy to check that the solution 
is the maximally correlated state
\begin{equation}\label{maxentstate}
|\Psi_0 \rangle =\frac 1{\sqrt{2J+1}}\sum_{M=-J}^J|M,-M\rangle
\end{equation}
Note that $|\Psi_0 \rangle$ also satisfies 
$(J_{1z}+J_{2z})|\Psi_0 \rangle =0$, and in fact, all spin components
have vanishing mean values in this state.

When we simulate the detection of phase shifts proportional to
$J_{1z}+J_{2z}$ and $J_{1y}-J_{2y}$, or combinations of these
operators, there is a chance, that the state vector 
is gradually projected onto $|\Psi_0 \rangle$, and this state
is unaffected by all future photodetection events. Indeed, in
some of our simulation records, we precisely see the robust
generation of the maximally entangled state, cf., Fig.~\ref{ent2}(a). 

If one has not collapsed into $|\Psi_0 \rangle$ after a large number of
photons have been detected, the state vector is instead orthogonal
to that state, and one will never arrive at the maximally entangled
state.  
The insets of the figures show the wave function overlap with $|\Psi_0
\rangle$, and in all simulations this quantity converges to either
unity, as in Fig.~\ref{ent2}(a), or to zero, as  in Fig.~\ref{ent2}(b). The overlap between
$|\Psi_0 \rangle$ and our initial state suggests that in one out of
$N+1$ realizations of the experiment, one should produce that particular
state, and this is confirmed by our simulations. 

There are no other joint eigenstates of the pair of collective
operators, and hence state vectors orthogonal to $|\Psi_0 \rangle$
do not evolve into any specific state: measurements on the system
have different outputs which affect the state vector
in different ways.  We do
note, however, that fairly strong entanglement is observed in many
realizations, and that this entanglement is almost constant over time. 
This is suggestive of families of states with relatively well defined 
values of the operators $J_{1z}+J_{2z}$ and $J_{1y}-J_{2y}$, which is
allowed by Heisenberg's uncertainty relation
as long as the expectation value of the operator $J_{1x}-J_{2x}$
is small. Figs. 6-8 present the three mean values of 
$J_{1x}-J_{2x}$, $J_{1y}-J_{2y}$ and $J_{1z}+J_{2z}$ for the 
evolution leading to the maximally entangled state, a very entangled
state and a less entangled state studied in Fig.~\ref{ent2} (a-c).
The picture is precisely as expected: the maximally entangled state
does not change with time and the mean values vanish forever,
a strongly entangled non-stationary state performs almost regular
oscillations in a limited part of Hilbert space restricting the mean
values to a similar small oscillatory behaviour, and states with little
entanglement show more dramatic time dependence and the mean values 
explore a wide range of values. It was unexpected that the non-maximally 
entangled states seem to be consistently much entangled or little 
entangled for long detection sequences. Further studies of this dynamics
would be very interesting.
       
\begin{figure}[hbtp]
\begin{center}
\includegraphics[width=8.5cm,]{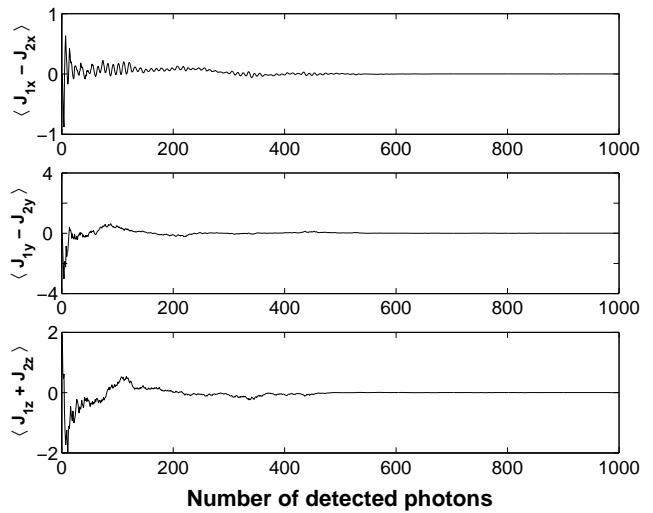}
\end{center}
\caption{Evolution of $\langle J_{1x}-J_{2x} \rangle$, $\langle
J_{1y}-J_{2y} \rangle$ and $\langle J_{1z}+J_{2z} \rangle$ corresponding to
the simulation shown in {\upshape Fig. \ref{ent2}(a)}. 
}
\label{a_mean}
\end{figure}

\begin{figure}[hbtp]
\begin{center}
\includegraphics[width=8.5cm,]{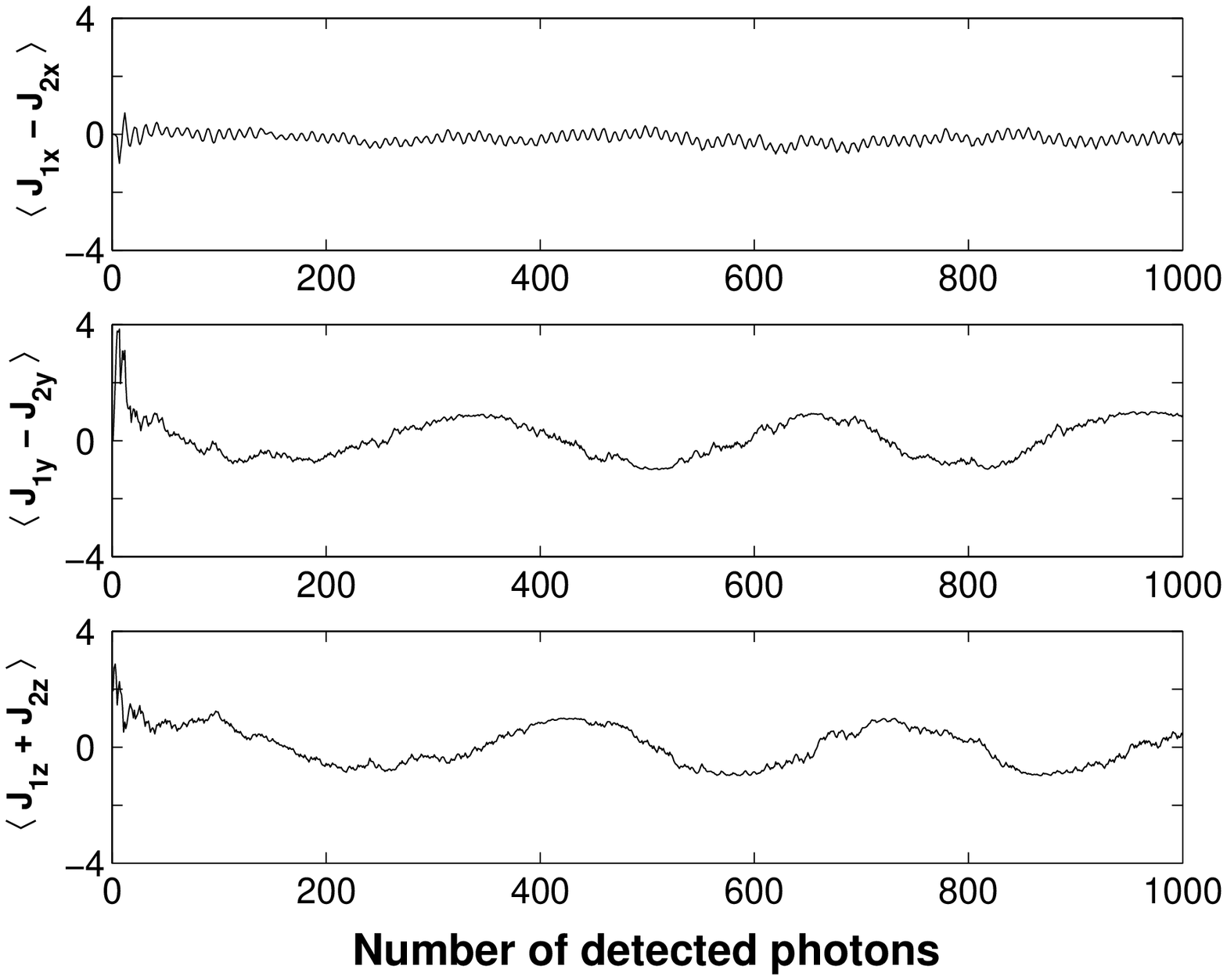}
\end{center}
\caption{Evolution of $\langle J_{1x}-J_{2x} \rangle$, $\langle
J_{1y}-J_{2y} \rangle$ and $\langle J_{1z}+J_{2z} \rangle$ corresponding to
the simulation shown in {\upshape Fig. \ref{ent2}(b)}. 
}
\label{b_mean} 
\end{figure}

\begin{figure}[hb]
\begin{center}
\includegraphics[width=8.5cm,]{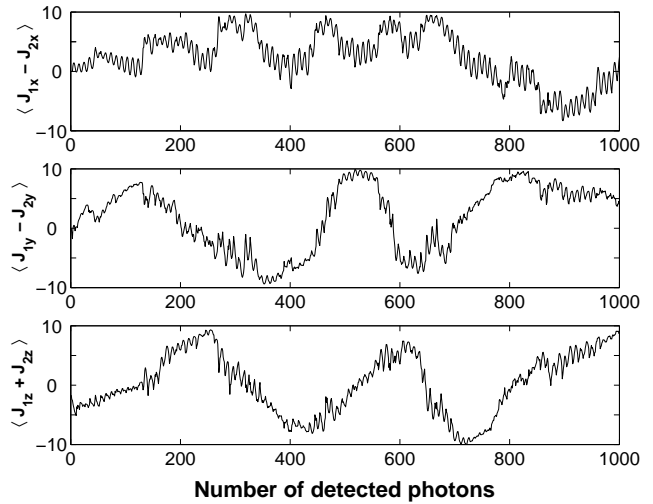}
\end{center}
\caption{Evolution of $\langle J_{1x}-J_{2x} \rangle$, $\langle
J_{1y}-J_{2y} \rangle$ and $\langle J_{1z}+J_{2z} \rangle$ corresponding to
the simulation shown in {\upshape Fig. \ref{ent2}(c)}. 
}
\label{c_mean} 
\end{figure}

\section{Conclusion}\label{conclusion}

We have presented a quantitative wave function analysis of the 
entanglement created by total population measurements on
separate atomic samples by means of optical phase shifts.
The analysis included wave function simulations where small samples
were exposed to interaction with the optical fields
in an interferometric set-up. We recall that more realistic experimental 
set-ups can be made , but they will be treated by the same formalism
and lead to the same results.  Our choice of
parameters (only 20 atoms, high atom-field coupling) was not meant to 
correspond to a specific experiment, but it brings out results, that we
may translate also to realistic experimental regimes. Note, however,
that the present work should be extended to include also spontaneous
emission by the atoms, if one wants to model experiments, where this
has significant effects on the dynamics.

The entanglement protocol has already been implemented experimentally
\cite{juls01}, and a theoretical analysis under an harmonic oscillator
approximation has been presented. On one side our analysis supplements
this existing analysis with a state vector perspective, and on the other
hand it provides an analysis valid beyond the oscillator approximation,
where in particular the emergence of a single maximally entangled state,
and families of non-stationary states with different degrees of
entanglement were identified. Since the experimentalist in principle
knows the state vector conditioned on the outcome of the 
detection, it seems interesting to introduce feed-back \cite{thomsen02}, either 
by just resetting the system and start over again if an only
weakly entangled state is prepared, or by suitably hitting the system
during measurements.

\end{document}